\documentclass[12pt]{article}
\usepackage{epsfig}
\usepackage{hhline}

\begin{document}

\title{Multiscale self-organized criticality and powerful X-ray flares}

\author{A. Bershadskii$^{1,2}$ and K.R. Sreenivasan$^2$}

\maketitle
\begin{center}
1.   {\it ICAR, P.O. Box 31155, Jerusalem 91000, Israel}

2.   {\it ICTP, Strada Costiera 11, I-34100 Trieste, Italy}
\end{center}

%%%%%%%%%%%%%%%%%%%%%%%%%%%%%%%%%%%%%%%%%%%%%%%%%%%%%%%%%%%%%%%%%%%%%%
\begin{abstract}
A combination of spectral and moments analysis of the continuous
X-ray flux data is used to show consistency of statistical
properties of the powerful solar flares with 2D BTW prototype
model of self-organized criticality.

\end{abstract}
%%%%%%%%%%%%%%%%%%%%%%%%%%%%%%%%%%%%%%%%%%%%%%%%%%%%%%%%%%%%%%%%%%%%%%


\newpage
\section{Introduction}

The solar flares are the manifestation of a sudden, intense and
spatially concentrated release of energy in the solar corona.
According to the Parker's idea \cite{parker} stochastic
photospheric fluid motions shuffle the footpoints of magnetic
coronal loops. The high electrical conductivity of the coronal gas
implies that the magnetic field is frozen-in, so that the
subsequent dynamical relaxation within the loop results in a
complex, tangled magnetic field, essentially force-free everywhere
except in numerous small electrical sheets which form
spontaneously in highly-stressed regions. As the current within
these sheets is driven beyond some threshold, reconnection sets in
and releases magnetic energy, leading to localized heating. In
this picture there exists a separation of timescales between
energy input to the system (minutes to hours, for photospheric
motions), and energy release (seconds to minutes, for reconnection
and thermalization under coronal conditions). It means that the
coronal loop is slowly driven by footpoint motions. Thus, the
coronal field does meet the requirements for the appearance of
self-organized criticality (SOC): self-stabilizing local threshold
instability; open boundaries; and separation of time scales
between driving and avalanching. Following to this picture the
authors of Ref. \cite{lh} suggested that SOC could describe the
main features of the hard X-ray flares with avalanches of the
reconnection events as the flares (see for further development
\cite{char}-\cite{hug} and references therein). Power law
probability density functions (PDF) observed for peaks, waiting
and duration times of the X-ray flares present the main
experimental support of this hypothesis. This experimental
information, however, turned out to be non-sufficient to
distinguish between the SOC and fluid (MHD) turbulence processes
mixed in the solar corona \cite{boff}. There exist substantial
differences between SOC and turbulence mechanisms for explanation
of the power law PDF's \cite{boff}. Such differences can be
ascribed to the lack of long time correlations for the avalanches
in the monoscale SOC models. The distinction between the two
mechanisms is however lost in the 2D BTW (Bak-Tang-Wiesenfeld)
model \cite{btw},\cite{dhar} which obeys a peculiar form of
multiscaling for the PDF's of several avalanche measures
\cite{mst}-\cite{ms}. The 2D BTW model has long time correlated
bursts and other turbulence-like intermittent properties if
studied at the wave time scales. It is suggested in \cite{ms} that
such SOC models would be candidates to describe the solar flares.
It should be also noted, that recent investigations
\cite{dp},\cite{snc} show that the lack of exponential
distribution for the waiting times between the solar flares (which
is mentioned in Ref. \cite{boff}) cannot be considered as a reason
for discarding the SOC models as underlying dynamics. The
long-range correlations of different nature can result in the
non-Poissonian behavior of the waiting times for the SOC systems.
The long-range correlations for the solar flares dynamics can have
both internal and external reasons (a correlated drive related,
for instance, to planetary motions affecting the Sun dynamics
\cite{snc}). The problem now is: How can one distinguish between
fluid (MHD) turbulence and the SOC models with long-time
correlations in the observed flares? We consider {\it continuous}
X-ray solar flux measured during a remarkable two-week period from
9 July to 23 July 2000 year. This period can be called as a flares
festival: 98 C-class, 39 M-class and 3 X-class flares. For
example, in analogous period of 1997 year the only one low-power
B-class flare was observed. In the minute time scale we have more
than 21000 data points presumably dominated by the powerful X-ray
flares. We will use a combined spectral-moments analysis of the
signal to extract information about underlying physical
mechanisms.

\section{Data and analysis}

The data was obtained by GOES-10 satellite of the National Oceanic
and Atmospheric Administration USA (NOAA) providing the kind of
continuous monitoring of the whole-sun X-ray flux, $I$, for the
0.5-to-4 A wavelenght band (1-minute averages). The data are given
in $Watts/cm^2~sec$. \\

We believe that a mix of different complex mechanisms produce the
total X-ray flux. Therefore, characteristics of the {\it raw}
signal generally do not exhibit some recognizable properties,
which could be used for identification of the underlying physical
processes. One should use some "threshold" technic to obtain the
recognizable features. On the other hand, even if such features
will appear, then a single certain characteristic, such as
spectrum or PDF, cannot univocally indicate a dominating
mechanism. For instance, different mechanisms (including SOC and
fluid turbulence) can produce the same power-law spectrum "-1". To
resolve this problem we will analyze two different physically
meaning fields: the flux, $I$, itself and its so-called local
dissipation rate \cite{ms},\cite{b}:
$$
\epsilon_t = \sum_{k=1}^t (I_{k+1}-I_k)^2/t  \eqno{(1)}
$$
Statistical properties of such gradient measure are generally
independent on statistical properties of the original time series
$I_k$. This fact allows us considering simultaneously the two
statistically independent fields: $I$ and $\epsilon =(\Delta
I)^2$, produce a reliable identification of the underlying
physical mechanisms. The name of the $\epsilon$ measure comes from
fluid turbulence, where the measure indeed is directly related to
the dissipation processes \cite{my}. Moreover, in our case the
$\epsilon$ field seems to be also directly related to dissipation
(see Introduction). Therefore, we will perform the threshold
discrimination of the data set using just the dissipation field
$\epsilon_k = (I_{k+1}-I_k)^2$. Then, we return to the $I$-field
to look at spectrum of $I$ after excluding the data points with
the large (above the threshold) dissipation. The recognizable
power-law spectrum of $I$ appears at low-frequencies at the
threshold value of the dissipation field $\epsilon \simeq
10^{-12}~W^2/cm^4 s^2$. The spectrum is shown in figure 2 (figure
1 shows the low-frequency spectrum of the {\it raw} flux $I$ for
comparison).

\begin{figure}[ht]
\epsfig{file=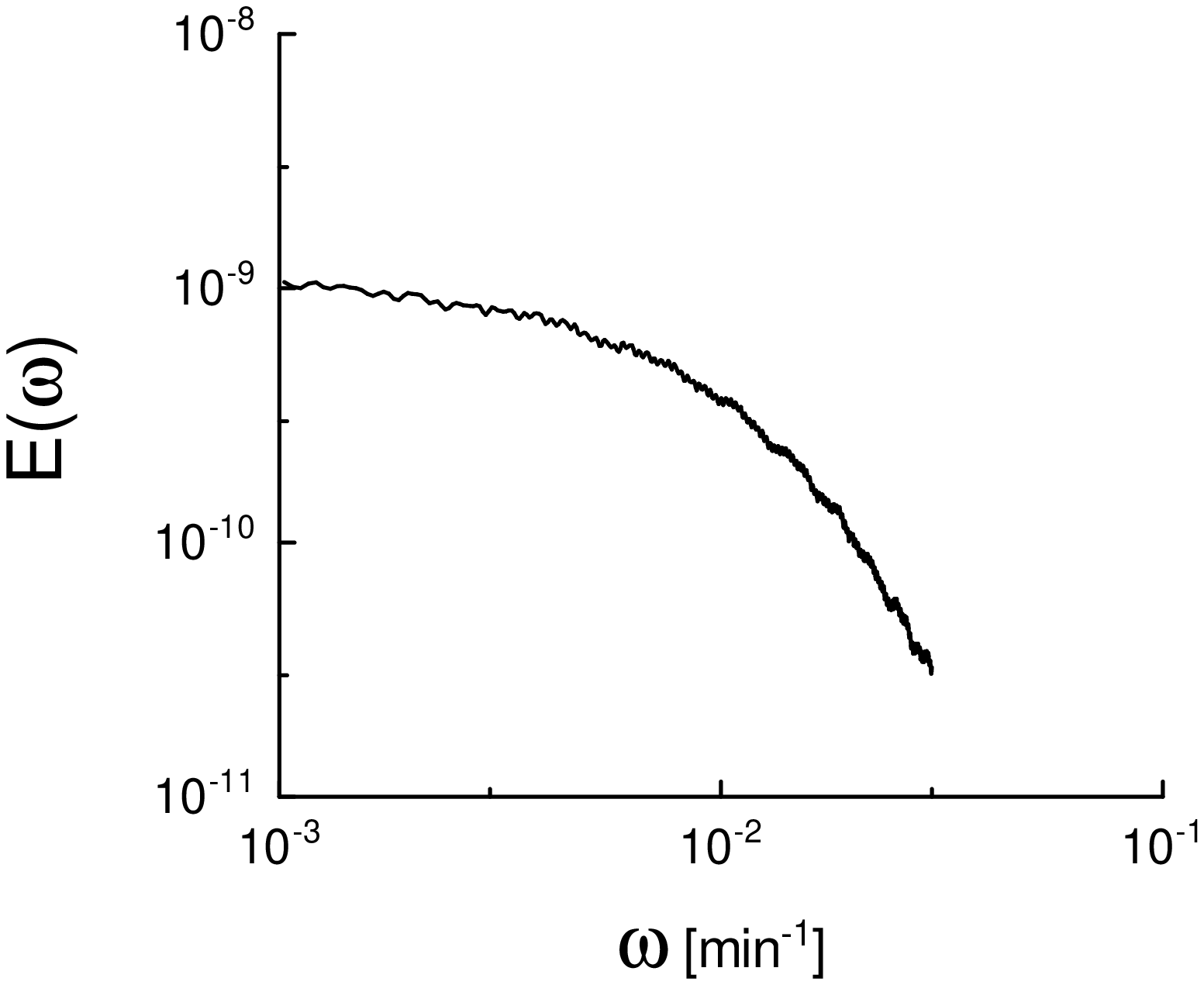,width=4.5in} \caption{\footnotesize Spectrum
of the X-ray flux raw data $I$ (in log-log scales).}

\end{figure}

\begin{figure}[ht]
\epsfig{file=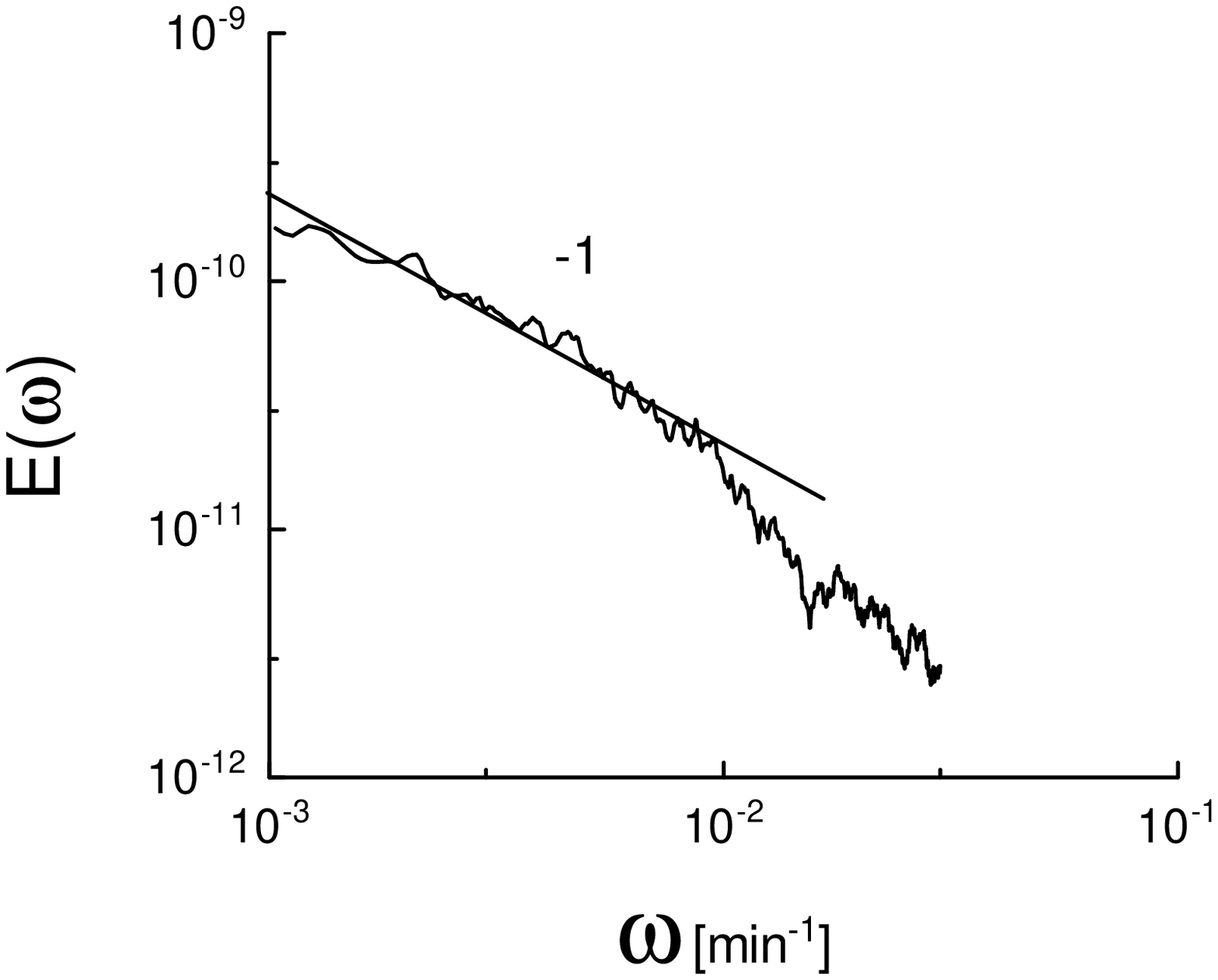,width=4.5in} \caption{\footnotesize Spectrum
of the X-ray flux data set, $I$, after exclusion of the points
with the local dissipation rate $\epsilon > 10^{-12}~ W^2/cm^4
s^2$. The straight line is drawn to indicate the "-1" power law
(2).}

\end{figure}

 The straight line is drawn in figure 2 in order to indicate the power-law
spectrum
$$
E(\omega) \sim \omega^{-1}   \eqno{(2)}
$$
in the log-log scales. For the discriminated data set (with the
spectrum (2) at the low frequencies) we calculated moments of the
local dissipation rate (1): $\langle \epsilon_t^p \rangle$. For
the multiscale SOC (in particular for the 2D BTW model \cite{ms})
appearance of the multiscaling
$$
\langle \epsilon_t^p \rangle \sim t^{-\mu_p}  \eqno{(3)}
$$
is expected. Figure 3 shows the moments against $t$ (in log-log
scales) and the straight lines are drawn in the figure in order to
indicate the multiscaling (3). Slopes of these straight lines
(best fit) provide us with multiscaling exponents $\mu_p$, which
are shown in figure 4 as circles. The time interval where we
observe the multiscaling of the dissipation field is: $10~min < Tt
< 1000~min$.

\begin{figure}[ht]
\epsfig{file=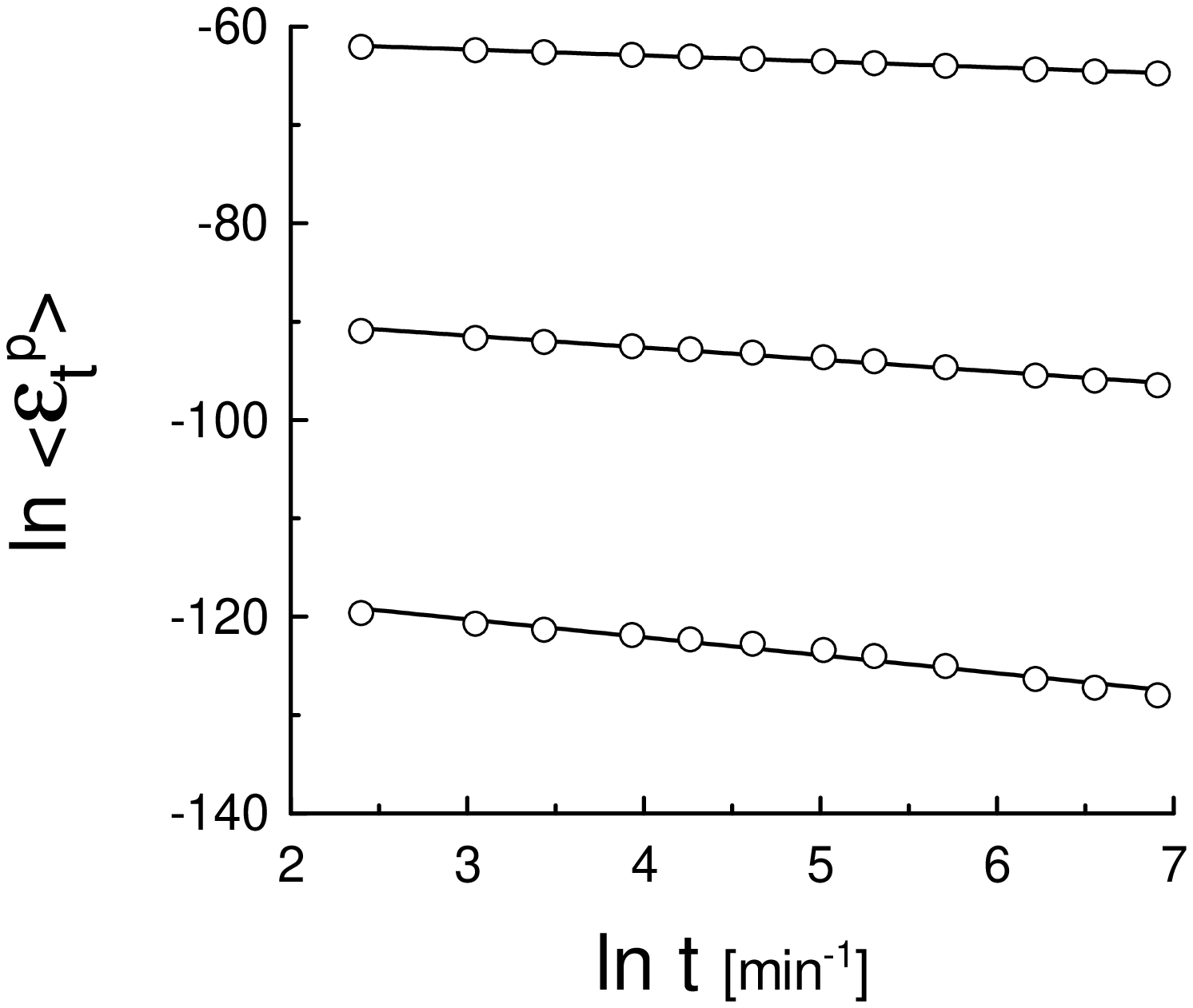,width=4.5in} \caption{\footnotesize The
local dissipation rate moments ($p=2,3,4$) against $t$ in log-log
scales. The data set is the same that was used to obtain figure 2.
The straight lines (the best fit) are drawn to indicate the
multiscaling (3).}

\end{figure}

\begin{figure}[ht]
\epsfig{file=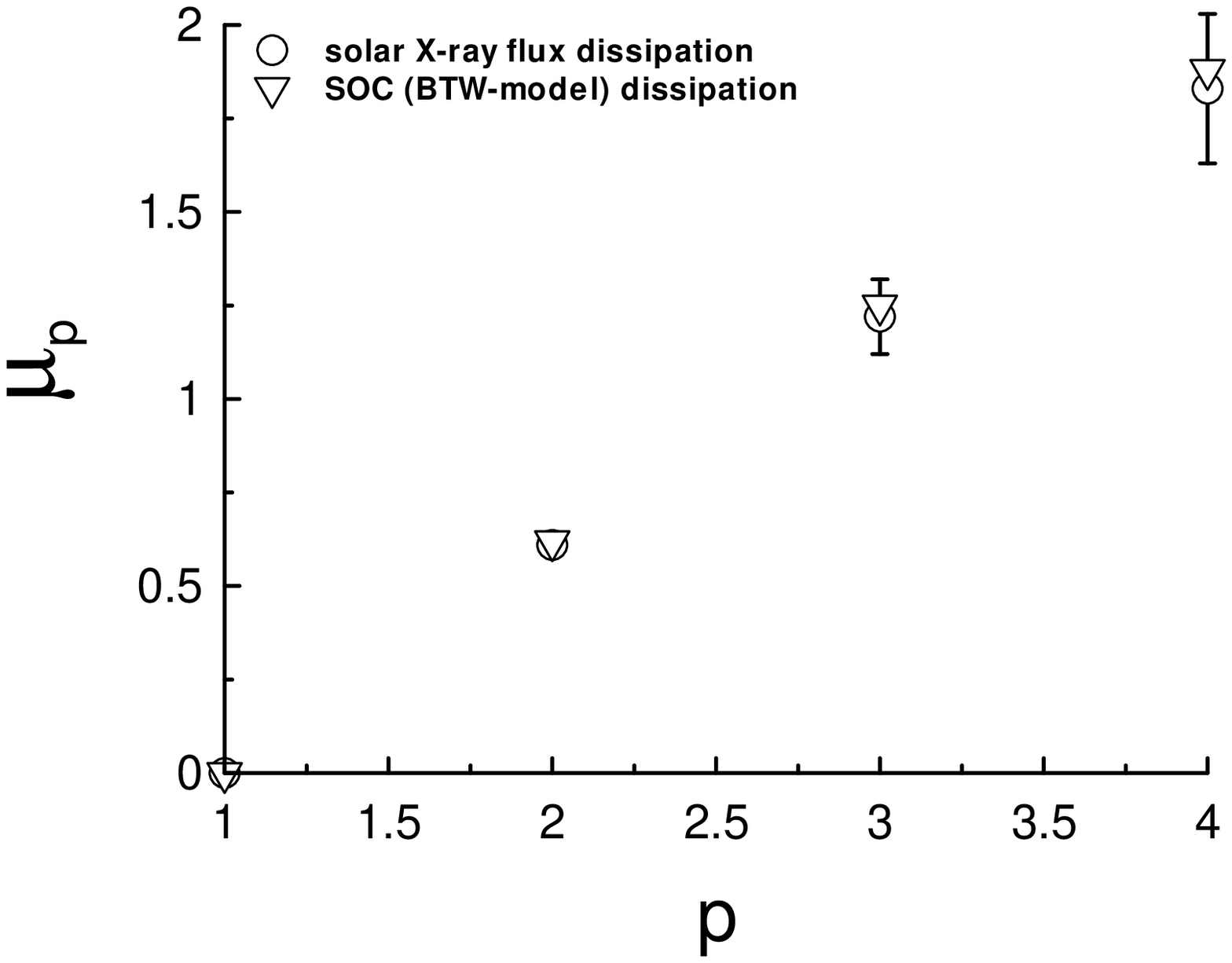,width=4.5in} \caption{\footnotesize The
multiscaling exponents $\mu_p$ (3) extracted as slopes of the
straight lines in figure 3 (circles). Triangles correspond to the
multiscaling exponents calculated in \cite{ms} for the 2D BTW
model.}

\end{figure}

\section{Discussion}

The BTW prototype model of self-organized criticality is defined
on a square lattice ($L\times L$) box \cite{dhar}.  On site $i$
$z_i=0,1,2,\ldots$ is the number of ``grains''. If $z_i<4$,
$\forall i$, the configuration is stable. Grain addition to a
stable configuration is made by selecting at random a site $k$
where $z_k \to z_k+1$. If then $z_k\ge 4$, toppling occurs, i.e.
$z_k \to z_k-4$, while each nearest neighbor, $l$, of site $k$
gets one grain ($z_l \to z_l +1$). If $k$ is at the border grains
are dissipated. The toppling of site $k$ may cause instabilities
in the neighbors, leading to further topplings at the next
microscopic time step, and so on. Thus, an avalanche made by a
total number $s\geq 0$ of topplings occurs before a new stable
configuration is reached and a new grain is added. After many
additions the system reaches a stationary critical state in which
avalanche properties are sampled.

A key notion in the approach to the BTW is that of wave
decomposition of avalanches \cite{ikp}. The first wave is obtained
as the set of all topplings which can take place as long as the
site of addition is prevented from a possible second toppling. The
second wave is constituted by the topplings occurring after the
second toppling of the addition site takes place and before a
third one is allowed, and so on.  The total number of topplings in
an avalanche is the sum of those of all its waves.  The wave size
has a pdf satisfying finite size scaling $P_w(s,L)\sim s^{-\tau_w}
f_w(s/{L^{D_w}})$, where $\tau_w=1$, $D_w=2$, and $f_w$ is a
suitable scaling function.

The multiscaling of the dissipation field
$$
\epsilon_t=\sum_{k=1}^t (s_{k+1}-s_k)^2 /t
$$
was observed in the Ref. \cite{ms} for the 2D BTW model and the
multiscaling exponents for the BTW dissipation calculated in
\cite{ms} are given in figure 4 as triangles. One can see good
correspondence between the exponents calculated in \cite{ms} and
the observed exponents.

The dissipation threshold $\epsilon \simeq 10^{-12}~ W^2/cm^4s^2$,
for which the solar X-ray data agree with the 2D BTW model both
for the flux spectrum and for the dissipation multiscaling,
corresponds to transition from the low intensity M-class flares to
the high intensity M-class flares. At present time we do not know
why just at this level of the dissipation rate the SOC (presumably
BTW) mechanism overcomes all other mechanisms and appears so clear
in the X-ray flux data. In any way, we can see that the multiscale
SOC takes place there and even appears as a dominating mechanism
at a certain stage.\\

The authors are grateful to the NOAA for providing the data.

\end{document}